\documentclass[12pt,a4paper,final]{iopart}

\usepackage{iopams}  
\usepackage{graphicx}
\usepackage[breaklinks=true,colorlinks=true,linkcolor=blue,urlcolor=blue,citecolor=blue]{hyperref}

\usepackage{dsfont}
\expandafter\let\csname equation*\endcsname\relax
\expandafter\let\csname endequation*\endcsname\relax

\usepackage{amsmath}
\usepackage{graphicx}
\usepackage{multirow}
\usepackage[latin1]{inputenc}
\usepackage{ulem} 
\usepackage{units} 
\usepackage{version} 
\usepackage{color}
\usepackage{colortbl}
\usepackage{xcolor}  

\newcommand\stxt[1]{_{\text{#1}}} 

\newcommand{\beq}{\begin{equation}}
\newcommand{\eeq}{\end{equation}}

\newcommand{\npeak}{n_0} 
\newcommand{\hatphik}{\hat{\phi}_k}
\newcommand{\hatnk}{\hat{n}_k}

\begin{document}




\title{Local relaxation and light-cone-like propagation of correlations in a trapped
one-dimensional Bose gas}

\author{R.~Geiger$^{1,*}$, T.~Langen$^1$, I.~E.~Mazets$^{1,2}$, J.~Schmiedmayer$^1$}
\eads{\mailto{remi.geiger@obspm.fr}}
\address{$^1$Vienna Center for Quantum Science and Technology, Atominstitut, TU Wien, Stadionallee 2, 1020 Vienna, Austria}
\address{$^2$Ioffe Physico-Technical Institute of the Russian Academy of Sciences, 194021 St.Petersburg, Russia }
\address{$*$ Present address : SYRTE, Observatoire de Paris, 61 avenue de l'Observatoire, 75014 Paris, France.}


\date{\today}

\begin{abstract}
We  describe the relaxation dynamics of a coherently split one-dimensional (1D) Bose gas in the harmonic approximation. 
A dephased, prethermalized state emerges in a light-cone-like evolution which is connected to the spreading of correlations with a characteristic velocity.
In our description we put special emphasis on the influence of the longitudinal trapping potential and the finite size of the system, both of which are highly relevant in experiments. 
In particular, we quantify their influence on the phase correlation properties and the characteristic velocity with which the prethermalized state is established.
Finally, we show that the trapping potential has an important effect on the recurrences of coherence which are  expected to appear in a finite size system.
\end{abstract}

\tableofcontents

\section{Introduction}
\label{par:intro}
The non-equilibrium dynamics of isolated quantum many-body systems has emerged as one of the recent frontiers in physics  \cite{CazalillaRigol2010NJP,Polkovnikov2011}. The question of how and under which conditions such systems may relax towards stationary states, as well as the applicability of statistical mechanics to describe these states has been of theoretical interest for several decades \cite{Neumann1929,Srednicki1994,Rigol2008ETH}. 
Important and yet unsolved  questions are, in particular, under which circumstances and on which time-scales many-body systems relax towards stationary states and whether these states can be described by statistical ensembles \cite{Srednicki1994,Berges2004,Rigol2007PRL,Rigol2008ETH,Berges2008,Eckstein2009,Kollar2011,Nowak2013}. 


Recently it has become experimentally possible to address these questions   with ultracold atomic gases \cite{Bloch2008,Cazalilla2011RevModPhys}.
Experiments with optically or magnetically trapped ultracold atoms which are well isolated from the environment represent good candidates for tackling difficult problems of quantum field theory, as they allow to tune several parameters (dimensionality, interactions, temperature, ...) and to probe many-body states in detail (see, e.g. \cite{Gring2012,Gerving2012,Cheneau2012,Hung2013}).
In ultracold atom experiments, the trapping potential is an essential ingredient, which is known to strongly affect the thermodynamics of the system \cite{Sagi2012,Schmidutz2013}. 
To compare many-body calculations with experiments it is thus desirable to study also the effect of the trapping potential on the non-equilibrium dynamics. 

In this work, we investigate the dynamics of a one-dimensional (1D) ultracold gas of bosons which is coherently split  along the transverse direction, as in the experiments presented in \cite{Gring2012,Kuhnert2013,Langen2013}. 
Previous theoretical studies have described the relaxation of the average coherence between the two gases \cite{Bistritzer2007}   as well as of the full distribution functions of interference contrast \cite{Kitagawa2011}, and addressed the questions of  dephasing beyond the harmonic approximation \cite{Burkov2007,Stimming2011,Mazets2009}. However, the influence of the trapping potential on the dynamics has not been discussed so far.
Here we develop a model for the dynamics of coherence between the two gases in the presence of a trap. We use this model to describe the evolution of local correlation functions, the spreading of dephasing, and the recurrences of coherence expected in finite size systems.

The paper is organized as follows : in the first part, we review the description of the homogeneous system in the phononic approximation and show how the dephased state emerges locally and spreads in space with the speed of sound. In the second part, we consider the trapped system and derive the expression of the two-point relative-phase correlation function between the two parts of the split system. Our model is in agreement with the experiments described in \cite{Langen2013}. 
Finally, we discuss further implications of the trap on the recurrences of coherence that are expected to appear in a finite size system as well as the  relevance of these recurrences for investigations of the dynamics beyond the harmonic approximation.


\section{Model}
We consider the dynamics of a gas of bosons that is confined in a radially symmetric harmonic trapping potential, with radial and longitudinal oscillation frequencies $\omega_\perp$ and $\omega$ respectively. The gas is in the quasi-1D regime \cite{Petrov2000} where the temperature $k\stxt{B}T$ and the chemical potential $\mu$ are much less than $\hbar\omega_\perp$, and $\omega\ll\omega_\perp$. The gas is coherently split symmetrically along the radial direction (for experiments, see e.g. \cite{Schumm2005,Gring2012}). The state of the system after splitting consists of two copies of the initial phase-fluctuating \cite{Petrov2000} condensate with (on average) $N/2$ atoms, where $N$ is the number of atoms in the gas before splitting. 
We assume that there is no tunnel coupling between the gases after splitting.

The Hamiltonian of the problem is given by :
\begin{equation}
\hat{\mathcal{H}} =  \int dz \sum_{j=1}^2  \Bigl(\frac{\hbar^2}{2m}\frac{\partial \hat\Psi_j^\dag}{\partial z}\frac{\partial\hat\Psi_j}{\partial z}  
      +\frac{g}{2}\hat\Psi_j^\dag\hat\Psi_j^\dag\hat\Psi_j\hat\Psi_j +(V(z)-\mu)\hat\Psi_j^\dag\hat\Psi_j \Bigr)             
   \label{eq:hamiltonian}
\end{equation}
with $\hat\Psi_j$   the atomic field operator associated to each gas ($j=1,2$),  $g=2\hbar\omega_\perp a$ is the 1D interaction strength~\cite{Olshanii1998}, $m$ is the atomic mass and $a$ is the 3D scattering length. For a homogeneous system in the Thomas-Fermi regime, the chemical potential is given by  $\mu=g \npeak$, with $\npeak$ being the linear density in each gas~\cite{Petrov2000}.
We decompose the field operators into $\hat\Psi_{j}=\sqrt{\hat{n}_{j}(z)}\exp[{i\hat{\theta}_{j}(z)}]$, with $\hat \theta_{j} (z)$ and $\hat n_{j} (z)$ denoting the operators describing the phase and density of each quasi-condensate, respectively. 
This approach allows to employ a generalization of standard Bogoliubov theory to describe the system ~\cite{Mora2003}, as density fluctuations in the two individual gases are suppressed and can be considered as small corrections. 
In the quasi-condensate regime (reduced density fluctuations), we have the commutation relation $[\hat{n}_i(z),\hat{\phi}_j(z^\prime)]=i\delta(z-z^\prime)\delta_{ij}$.

\section{Relaxation of the homogeneous Bose gas}
\label{par:homog}

\subsection{Hamiltonian and phononic excitations}
We start by studying the case of an homogeneous gas with $V(z)=0$.  We quickly recall the theoretical analysis introduced in \cite{Langen2013epjst} and derive  the time evolution of the two-point relative-phase correlation function (PCF) which can be used to characterize the  decay of coherence locally. The situation of two quasi-condensates in thermal equilibrium has been  addressed in \cite{Bouchoule2003}. Here we discuss the relaxation of the PCF after the coherent splitting described above and explain the local equilibration observed recently in the experiment of Ref. \cite{Langen2013}.

We introduce the relative phase and relative density between the two gases through $\hat{\phi}=\hat{\theta}_1-\hat{\theta}_2$ and $\hat{n}=(\hat{n}_1-\hat{n}_2)/2$, with the commutation relation $[\hat{n}(z),\hat{\phi}(z^\prime)]=i\delta(z-z^\prime)$. This is motivated by interference experiments which give direct access to the relative phase field.
The common (center-of-mass) degrees of freedom are described by the operators $\hat{\phi}_c=(\hat{\theta}_1+\hat{\theta}_2)/2$ and $\hat{n}_c=\hat{n}_1+\hat{n}_2$.
Inserting the definition of $\hat{\Psi}_j$ in eq. \eqref{eq:hamiltonian} and neglecting third order terms  ($\sim (\partial_z \hat{\theta})^2 \hat{n}$), we obtain a quadratic Hamiltonian for the phase and density operators. 
Assuming a symmetric splitting (equal density in each gas after splitting), the dynamics of the relative and the common degrees of freedom decouple, as shown in \cite{Kitagawa2011}. We obtain the  following  Hamiltonian governing the evolution of the relative phase and density operators:
\beq
\hat{H} = \int dz \Big[\frac{-\hbar^2 \npeak}{4m} (\partial_z \hat{\phi})^2 + g\hat{n}^2 \Big].
\eeq
We rewrite this Hamiltonian in the more usual form of a Luttinger liquid \cite{Giamarchi2004}
\beq
\hat{H} = \frac{\hbar c}{2}\int dz \Big[\frac{K}{\pi} (\partial_z \hat{\phi})^2 + \frac{\pi}{K}\hat{n}^2 \Big],
\label{eq:Luttinger}
\eeq
with the speed of sound $c=\sqrt{g\npeak/m}$, the Luttinger parameter $K=\frac{\hbar\pi}{2}\sqrt{\frac{\npeak}{mg}}$ and $\npeak$ the 1D peak density.
The Luttinger liquid model is the long wavelength approximation of the many-body Hamiltonian \eqref{eq:hamiltonian} and allows to describe the phononic (long wavelength) excitations of the system. 
This approximation is valid for the description of typical ultracold atom experiments, where optical methods with an imaging resolution corresponding to a few times the healing length are used to probe the excitations in the system.

Considering periodic boundary conditions for a system of size $\mathcal{L}$, we expand the field operators as \cite{Langen2013epjst}
\beq
 \hat{n}(z,t) = \frac{1}{\sqrt{\mathcal{L}}} \sum_{k\neq 0} \hatnk(t) e^{ikz}  \ \ , \  \hat{\phi}(z,t) = \frac{1}{\sqrt{\mathcal{L}}} \sum_{k\neq 0} \hatphik(t) e^{ikz}
\label{eq:expansion_homog}
\eeq
with the expansion coefficients given by:
\begin{eqnarray}
\hatnk(t) = \sqrt{\frac{\npeak S_k}{2}} \Big( \hat{b}_k(t)  + \hat{b}_{-k}^\dagger(t)  \Big)  \nonumber  \\
\hatphik(t) = \frac{1}{i\sqrt{2\npeak S_k}} \Big( \hat{b}_k(t)  - \hat{b}_{-k}^\dagger(t)  \Big).
\end{eqnarray}
Here   $\hat{b}_{k}^\dagger$ and $\hat{b}_k$ are the creation and annihilation  operators for an elementary excitation with momentum $\hbar k$ in the relative degrees of freedom ($k=p\times 2\pi/\mathcal{L}$ with $p$ integer different than $0$ ; note that the sum expands over both positive and negative $k$). 
The commutation relation for the expansion coefficients reads $[\hatnk,\hatphik^\dagger]=[\hatnk,\hat{\phi}_{-k}]=i$.
The structure factor in the phononic regime is given by
\beq
S_k = \frac{\hbar |k|}{2 m c} = \frac{|k| K}{\pi \npeak},
\label{eq:struct_factor}
\eeq
and is related to the $(u_k,v_k)$ Bogoliubov coefficients used in previous works \cite{Mora2003,Bouchoule2003} through $\sqrt{S_k} = (u_k-v_k)^{-1}$.

In this basis, the Hamiltonian \eqref{eq:Luttinger} takes the diagonal  form
\beq
\hat{H} = \frac{\hbar c}{2} \sum_{k\neq 0} \Big[ \frac{K}{\pi}k^2 \hatphik^\dagger \hatphik + \frac{\pi}{K} \hatnk^\dagger \hatnk \Big] + \frac{\hbar \pi c}{2 K} \hat{n}_0^\dagger\hat{n}_0 = \sum_{k\neq 0} \hbar \omega_k \hat{b}_{k}^\dagger \hat{b}_{k} + \frac{\hbar \pi c}{2 K} \hat{n}_0^\dagger\hat{n}_0,
\label{eq:Luttinger_k}
\eeq 
with $\omega_k=c|k|$. We have made explicit the term corresponding to the $k=0$ mode which accounts for the  global phase diffusion.
The dynamics can thus be reduced to that of a set of uncoupled harmonic oscillators. It follows from the absence of coupling between modes with different momentum $k$ that any momentum occupation numbers $\langle \hat{b}_{k}^\dagger \hat{b}_{k}\rangle$ that are initially imposed on the system will  be conserved. Eq. \eqref{eq:Luttinger_k} therefore strikingly shows the integrability of the system.

\subsection{Initial conditions and equations of motion.} 
We assume the coherent splitting process to be instantaneous (fast with respect to the interaction
energy, $t\stxt{split}\ll h/\mu=2\pi\xi_h/c$) so that it can be described like a beam splitter in optics.  Consequently the local distribution of atoms in each small region of the quasi-condensate (of size  $\sim \xi_h=\hbar/mc$) is binomial, with the respective minimum uncertainty relative-phase distribution.
In that way, the coherent splitting copies the phase fluctuations of the initial quasi-condensate into both parts of the split system and the  relative density fluctuations are given by the local shot noise. In terms of elementary excitations, this means : 
\beq
\langle\hatphik^\dagger\hatphik\rangle|_{t=0} = \frac{1}{2\xi_n^2\npeak} \ \ , \  \langle \hatnk^\dagger \hatnk \rangle|_{t=0} = \frac{\xi_n^2\npeak}{2}
\label{eq:init_conditions}
\eeq
with $\xi_n^2$ being the number squeezing parameter (in the following we will take $\xi_n^2=1$ unless specified). 
A state where the fluctuations are initialized with an excess of relative-density fluctuations and a lack of relative-phase fluctuations is a non-equilibrium state far away from the thermal equilibrium state of the Hamiltonian \eqref{eq:Luttinger_k} at a temperature $T$ where: 
\beq
\langle\hatphik^\dagger\hatphik\rangle|_{\text{th}} = \frac{2}{\lambda_T k^2} \ \ , \  \langle \hatnk^\dagger \hatnk \rangle|_{\text{th}} = \frac{k\stxt{B}T}{2g}
\label{eq:GS}
\eeq
with $\lambda_T=\hbar^2 \npeak /mk\stxt{B}T$.
The  atom number fluctuations caused by the splitting correspond to the introduction of energy into the system due to the interactions. This energy  manifests itself in additional excitations  in the density quadrature (rather than in the phase quadrature)  with respect to the thermal equilibrium (compare Eqs. \eqref{eq:init_conditions} \eqref{eq:GS}).

In terms of the elementary excitations, the initial conditions of Eq. \eqref{eq:init_conditions} lead to an approximately thermal-like form of the occupation numbers that reads
\beq
\langle \hat{b}_k^\dagger \hat{b}_k\rangle = \frac{k_BT\stxt{eff}}{\hbar\omega_k},
\label{eq:init_occu_nb}
\eeq
with the effective temperature given by $k_BT\stxt{eff}=\xi_n^2\npeak g/2$. The fast splitting process thus equally distributes the energy $k_BT\stxt{eff}$ in the different modes of the system \cite{Kitagawa2011}.

The equations of motions for the operators $\hatphik (t)$ and $\hatnk (t)$ read
\begin{eqnarray}
i\hbar \frac{d\hatphik}{dt} = [\hatphik,\hat{H}] = \frac{\hbar c }{2}\frac{\pi}{K} (-2i \hatnk) \nonumber  \\
i\hbar \frac{d\hatnk}{dt} = [\hatnk,\hat{H}] = \frac{\hbar c }{2} \frac{K}{\pi} k^2 (2i\hatphik) 
\label{eq:eq_motion}
\end{eqnarray}
and lead to  harmonic oscillator equations for $\hatnk$ and $\hatphik$, with the angular oscillation  frequency $\omega_k=ck$.
With the initial conditions \eqref{eq:init_conditions} and neglecting the initial phase fluctuations ($\langle\hatphik^\dagger\hatphik\rangle|_{t=0}\approx 0$, which is valid as soon as $t > h/\mu$), we obtain the evolution of the expansion coefficients:
\begin{eqnarray}
\hatphik(t) = \frac{\pi}{kK}\sqrt{\frac{\xi_n^ 2\npeak}{2}}\sin\omega_kt \nonumber \\
\hatnk(t) = \sqrt{\frac{\xi_n^ 2\npeak}{2}}\cos\omega_kt.
\label{eq:evolution}
\end{eqnarray}
These equations show the oscillation between the density and phase quadratures for the phononic excitations.

\subsection{Relative-phase correlation function.}
\label{par:corr_func}
To study the decay of coherence between the two quasi-condensates  and the evolution of correlations in the system, we consider the two point relative-phase correlation function defined as \cite{Langen2013epjst}: 
\begin{align}C(z,z',t)&= \frac{\langle \hat{\Psi}_1(z) \hat{\Psi}^\dagger_2(z) \hat{\Psi}^\dagger_1(z^\prime) \hat{\Psi}_2(z^\prime)\rangle}{\langle|\hat{\Psi}_1(z)|^2\rangle\langle|\hat{\Psi}_2(z^\prime)|^2\rangle}\nonumber\\
&\simeq\exp\left(-\frac{1}{2}\langle\Delta\phi_{zz'}(t)^2\rangle\right).
\label{eq:PCF}
\end{align}
In the last step we have used the fact that the fluctuations are Gaussian, which is a consequence of the quadratic Hamiltonian Eq. \eqref{eq:Luttinger}. 
Here, $\langle\Delta\phi_{zz'}(t)^2\rangle\equiv\langle(\hat{\phi}(z,t)-\hat{\phi}(z',t))^2\rangle$ denotes the phase variance between two points z and z' of the relative phase field. 
Using \eqref{eq:evolution}, the time evolution of the phase variance is given by 
\begin{equation}
	\langle\Delta\phi_{zz'}(t)^2\rangle = \frac{\pi^2 \npeak}{\mathcal{L} K^2} \sum_{k\neq0} \frac{\sin(\omega_k t)^2}{k^2}\left(1-\cos(k\bar z)\right).
\label{eq:phase_variance}
\end{equation}
The first term in the sum \eqref{eq:phase_variance} represents the growth and subsequent oscillations in the amplitude of the phase fluctuations as they get converted from the initial density fluctuations. The factor $1/k^2$ in the amplitude reflects the $1/k$ scaling of the excitation occupation numbers associated with the equipartition of energy induced by the fast splitting, Eq. \eqref{eq:init_occu_nb}. The  term in parenthesis in the sum corresponds to the spatial fluctuations.  

We will discuss the physical interpretation of this expression in more details in the next two paragraphs, considering first the dephased (prethermalized) state and second the dynamics governing the emergence of this state.

\subsection{Prethermalized state}

We first focus on the dephased state of the system where the two-point relative-phase variance  can be approximated by its long time limit, taking $\sin^2(\omega_kt)=1/2$. Approximating the sum by an integral, we  obtain 
\beq
\langle\overline{\Delta\phi_{zz'}^2}\rangle = \frac{4mg}{\hbar^2}\times \frac{1}{2} \times \frac{|\bar z|}{2},
\eeq  
where we have used $\int \frac{dk}{2\pi} \frac{1-\cos(k \bar z)}{k^2}=\frac{|\bar z|}{2}$.
In the relaxed (dephased) state, the phase correlation function is thus given by 
\beq
C(\bar z) = \exp\Big(-\frac{|\bar z|}{l_0}\Big), \ \text{with} \ l_0 = \frac{2\hbar^2}{mg}.
\eeq
The correlation length $l_0$ does not depend on the detail of the system (such as the density) but only on the 1D coupling constant, and  is in that sense universal. The scaling comes from  the equipartition of energy between the modes during the splitting.

\subsection{Multimode phase diffusion}
\label{par:multimode_or_monomode}
It is interesting to discuss the role of the correlation length $l_0$ in the context of dephasing, linking our study with previous results obtained on phase diffusion in 3D BEC~\cite{Castin1997,Javanainen1997,Leggett1998} and quasi-condensates~\cite{Bistritzer2007}.
The decay of the coherence factor $\Psi(t)\equiv\langle e^{i\hat{\phi}(z,t)} \rangle = e^{ - \frac{1}{2} \langle \hat{\phi}(z,t)^2 \rangle}$ due to the lowest mode ($k=0$ term in Eq. \eqref{eq:Luttinger_k}) is given by : 
\beq
\Psi(t)|_{k=0} \propto e^{-t^2/\tau_0^2}
\eeq
with the characteristic time $\tau_0$ known from the phase diffusion in 3D BEC~\cite{Castin1997,Javanainen1997,Leggett1998}: $\tau_0=\frac{\hbar}{g}\sqrt{\frac{L}{\npeak}}$.
On the other hand, the decay of coherence due to the contribution of all other ($k\neq 0$) modes is given by~\cite{Bistritzer2007}
\beq
\Psi(t)|_{k\neq 0} \propto e^{-t/\tau}
\eeq
with the dephasing time $\tau = 8K^2/\pi^2\npeak c$ (note that our definition of $K$ differs by $1/2$ with respect to \cite{Bistritzer2007}).
The dephasing will therefore be dominated by 1D effects as soon as $\tau<\tau_0$, which correspond to the condition
\beq
l_0 = \frac{1}{\xi_n^2} \frac{2\hbar^ 2}{mg} < \frac{L}{2},
\label{eq:condition_multimode}
\eeq 
where we  explicitly introduced the squeezing parameter.

\begin{figure}[h!]
	\centering
	\includegraphics[width=0.7\textwidth]{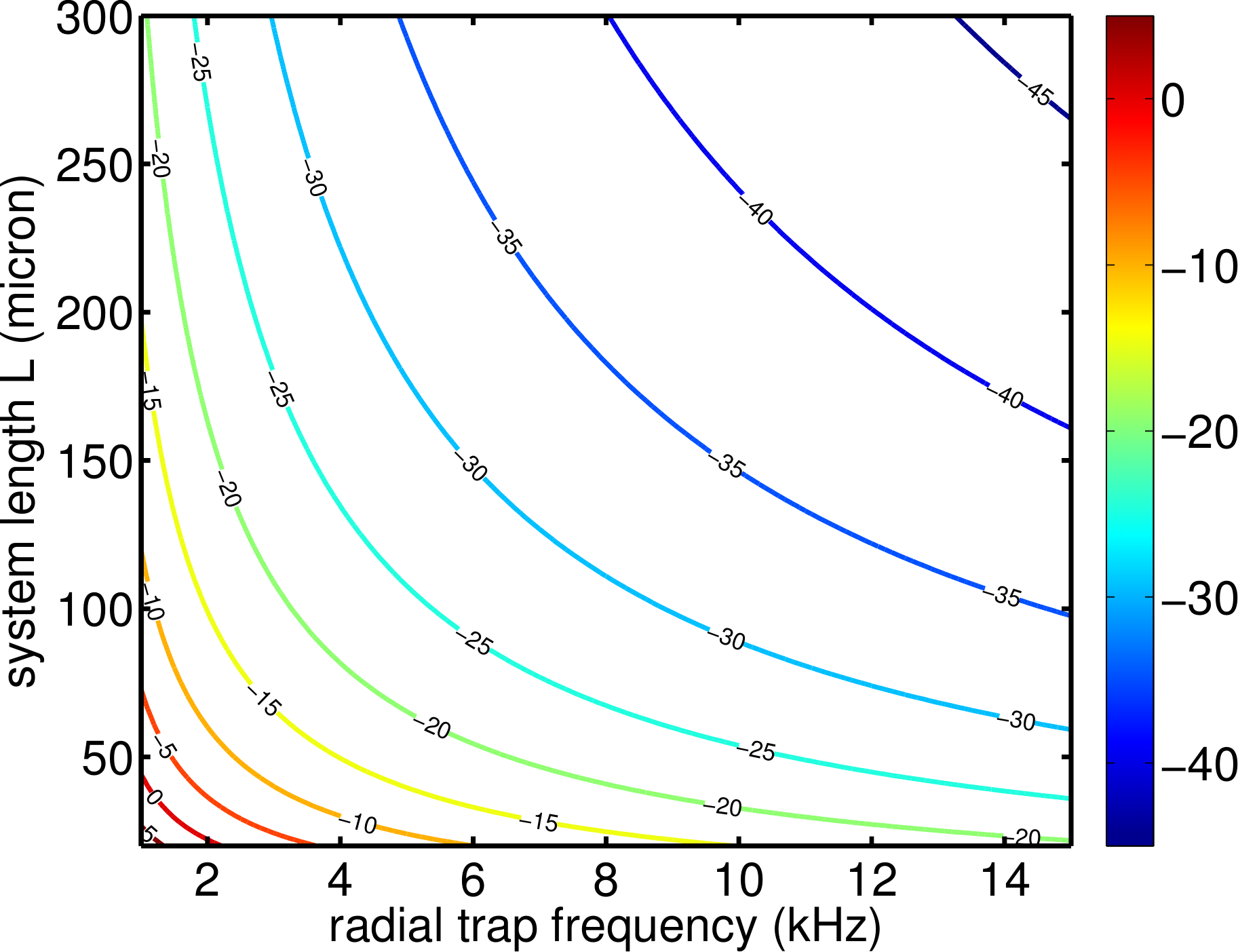} 
	\caption{Required number squeezing $\xi_n^2|_{\text{lim}}$ for an interferometer to be limited by 3D phase diffusion rather than 1D (multimode) effects. The figure shows a contour plot of the required squeezing in dB versus the system length $L$ and the radial trap frequency $\omega_\perp/2\pi$. 
	}
	\label{fig:squeezing_multimode}
\end{figure}

Recent studies on the role of atom-atom interactions in interferometers indicate that  low-dimensional geometries could be favourable for interferometry \cite{Grond2010}.  
However, 1D effects will limit this coherence time  if the prethermalized correlation length is smaller than half the system size, whereas they can be neglected with respect to the 3D phase diffusion if $l_0>L/2$. 
The latter is, for example, the case in experiments that are working with low atom numbers, high number squeezed states  or not too elongated traps \cite{Maussang2010,Berrada2013}.
Equation \eqref{eq:condition_multimode} can be re-formulated in terms of experimentally controllable parameters, by expressing the amount of number squeezing required for a beam splitter so that the interferometer will not be limited by multimode dephasing :
\beq
\xi_n^2<\xi_n^2|_{\text{lim}} = \frac{2\hbar}{m\omega_\perp a_s L}.
\label{eq:metrology}
\eeq 
We used $g=2\hbar\omega_\perp a_s$ (in the quasi-1D regime) to express the number squeezing parameter in terms of measurable quantities (radial trap frequency $\omega_\perp$ and system length $L$).
Equation \eqref{eq:metrology} provides a  criterion for evaluating the possible sensitivity of elongated condensates working close to the 1D regime for applications in interferometry. We illustrate this criterion in Fig.~\ref{fig:squeezing_multimode} where we represent the required number squeezing (in dB) for an interferometer to be limited by 3D phase diffusion rather than 1D effects, as a function of the physical parameters $(\omega_\perp, L)$ controlling the 1D-ness of the system. 
We finally note that in the homogeneous case, condition    \eqref{eq:metrology} does not involve the linear density of the system as the dephasing time due to the $k=0$ mode and that due to the $k \neq 0$ modes have the same dependence $\propto \npeak^{-1/2}$. In practical applications, however, the system size  and the peak density of the trapped atoms are mutually related, thus leading to an implicit density dependence of the condition \eqref{eq:metrology}.

\subsection{Local relaxation and light-cone effect}

\begin{figure}[h!]
	\centering
	\includegraphics[width=0.7\textwidth]{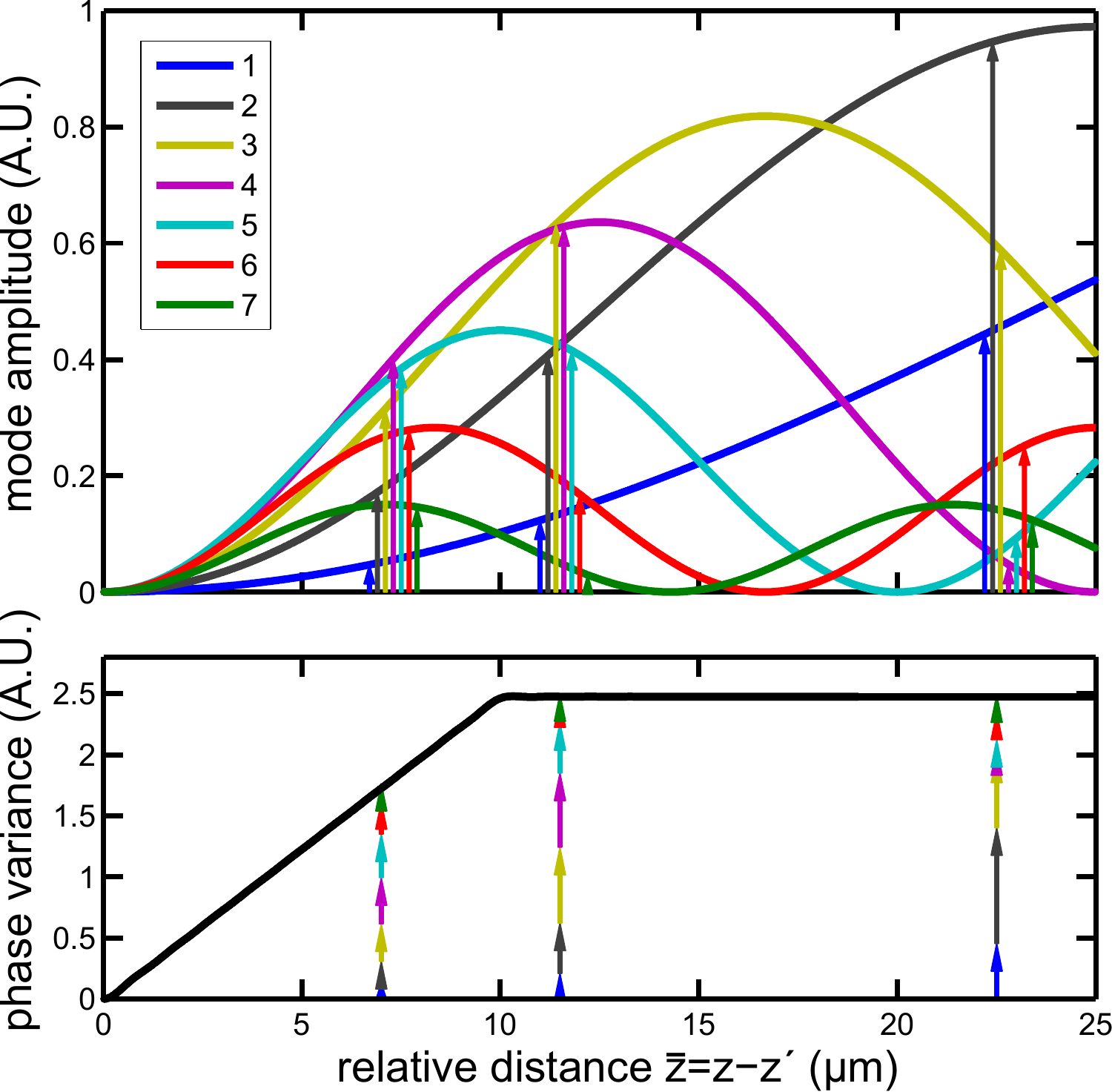} 
	\caption{Visualization of the light-cone-condition for $c=1\,$mm/s and $t=5\,$ms. The relative phase of the system is randomized by a superposition of many modes (solid lines, the first seven modes are represented). Initially, the contribution of all these modes grows in amplitude (arrows), leading to a linear increase in the variance of the phase (bottom plot). For the correlation function, this corresponds to the establishment of thermal correlations up to $\bar z_c=2ct$. Beyond $\bar z_c$, modes with a wavelength larger than $\bar z_c$ would be needed for a further randomization of the phase. However, while these long-wavelength modes grow in amplitude for $\bar z>\bar z_c$, modes with shorter wavelength start to decrease again in amplitude. Overall, this leads to a constant phase variance beyond $\bar z_c$. Figure adapted from \cite{Langen2013}.}
	\label{fig:illustration_modes}
\end{figure}

We now discuss the full expression Eq. \eqref{eq:phase_variance} which characterizes the decay of coherence  as a result of the superposition of many phononic modes.
Mathematically, expression \eqref{eq:phase_variance} is the Fourier decomposition of a trapezoid with a siding edge at $\bar z_c =2ct$, so that the phase variance (and thus the phase correlation function) exhibits a two step feature. 
This two-step feature can physically be interpreted as follows: for a given time $t$, short wavelength modes will grow in amplitude and linearly increase the phase variance up to a distance $\bar z_c=2ct$. Beyond that point the growth in amplitude of longer wavelength modes with $2\pi/k>\bar z_c$ exactly compensates the decrease in amplitude of the shorter wavelength modes with $2\pi/k<\bar z_c$, leading to a constant phase variance. This is illustrated in Fig. \ref{fig:illustration_modes}. 

\begin{figure}[h!]
	\centering
	\includegraphics[width=0.6\textwidth]{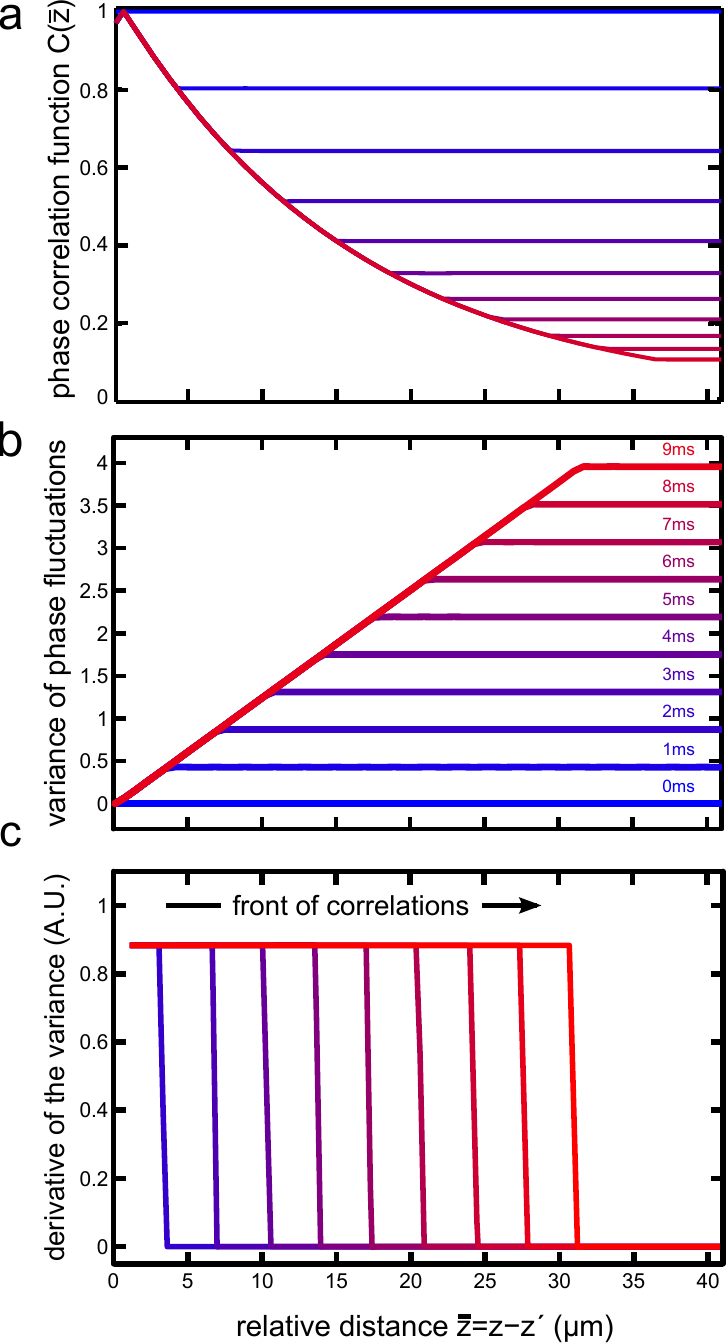} 
	\caption{Figure adapted from \cite{Langen2013}. (a) 	Time-evolution of the relative-phase correlation function for $c = 1.8 \ $mm/s.
	(b) Time evolution of the relative-phase variance $\langle\Delta\phi_{zz'}^2\rangle$. For a given evolution time, the phase variance grows linearly up to a distance $\bar z_c = 2 c t$. Beyond that sharp crossover point the phase variance is constant, revealing the persisting long-range phase coherence in the system. (c) Derivative of the phase variance visualizing how the front of correlations travels through the system. }
	\label{fig:time_evolution}
\end{figure}

Quantitatively, computing the derivative of the phase variance with respect to time, we find 
\begin{equation}
	\frac{\partial \langle\Delta\phi_{zz'}(t)^2\rangle}{\partial t} = \frac{2c}{l_0} \times \Theta(2ct-\bar z),
	\label{eq:deriv_phase_var}
\end{equation}
with $\Theta(x)$ being the Heaviside step function ($\Theta(x)=1$ if $x>0$ and 0 otherwise) and $l_0=2\hbar^2/mg$ the prethermalized correlation length introduced in Eq. \eqref{eq:condition_multimode}. The phase randomizes  with a constant rate up to the point where $\bar z = 2 c t$. Beyond that point, long-range phase coherence is retained. 
The full time evolution of the phase variance and its derivative, and the experimentally probed phase correlation function is shown in Fig.~ \ref{fig:time_evolution}.

The form of the phase correlation function reveals the physical interpretation of Eq. \eqref{eq:deriv_phase_var} :
correlations in the system take the final relaxed form (exponential decay with correlation length $l_0$) within the propagation cone, while long-range order remains outside of the cone. The separation $2ct$ corresponds to the distance traveled by quasi-particles moving with a velocity $c$ in opposite directions. These quasi-particles establish the  correlations between points $z$ and $z^\prime=\bar z +z $ when they meet at a time $t=\bar z/2c$ \cite{Calabrese2006}.
Such light-cone effects have been found in various quantum many-body systems \cite{Mitra2011,Barmettler2012,Mitra2013,Carleo2013arxiv,Deuar2013arxiv}, and the local relaxation picture was first introduced in  a general way in the context of the Bose-Hubbard model \cite{Cramer2008}.
Here, we provide a connection to the relaxation of a Luttinger liquid, finding a similar result as the one obtained numerically for the Bose-Hubbard model \cite{Cramer2008}. 



We finally note that while the squeezing parameter $\xi_n^2$ influences the final relaxed state (influence on the correlation length $l_0$), it  does not modify the velocity of correlations, i.e. the time after which the relaxed state is reached.





\section{Relaxation of the trapped system}
\label{par:trap}

\subsection{Relative-phase correlation function}

We now investigate the dynamics of coherence between the two gases in the case of a longitudinal trapping potential $V(z)=\frac{1}{2} m \omega^2 z^2$ (see Eq. \eqref{eq:hamiltonian}). Under the same approximations as for the homogeneous case (neglecting third order terms in the fluctuations and considering only phononic modes), the Hamiltonian for the relative degrees of freedom can be written as 
\beq
\hat{H} = \int dz \Big[\frac{-\hbar^2 n_0(z)}{4m} (\partial_z \hat{\phi})^2 + g\hat{n}^2 \Big],
\label{eq:hamiltonian_trap}
\eeq
where the density profile $n_0(z)$ solves the Gross-Pitaevski equation
\beq
\Big[ \frac{-\hbar^ 2}{2m} \Delta +(V(z)-\mu) +g n_0(z) \Big] \sqrt{n_0(z)} = 0.
\eeq 
The Hamiltonian thus takes the form of a non-homogeneous Luttinger Liquid with spatially dependent velocity $c(z) = \sqrt{n_0(z)g/m}$ and Luttinger parameter $K(z)=\frac{\hbar\pi}{2}\sqrt{\frac{n_0(z)}{mg}}$.

Following the notations of Petrov et al. \cite{Petrov2004}, this Hamiltonian can be re-written as
\beq
\hat{H} = \frac{\hbar}{2\pi} \int dz \Big[ v_N (\pi\hat{n})^2 + v_J (\partial_z \hat{\phi})^2 \Big],
\eeq
with the density and phase velocities given by $v_N=c/K$ and $v_J=cK$.

For a weakly interacting gas in the Thomas-Fermi regime, the density profile is an  inverted parabola
\beq
n_0(z) = \npeak \Big(1-\frac{z^ 2}{R^2} \Big),
\label{eq:TF}
\eeq
with the Thomas-Fermi radius $R=\sqrt{2}c_0/\omega$ given by peak density of the gas via $c_0=\sqrt{g\npeak/m}$ and the peak density determined from the atom number \cite{Petrov2000}. In the Thomas-Fermi regime, the velocity $v_N$  reduces to $v_N = 2g/\pi\hbar$ and is therefore independent on the spatial coordinate.

Proceeding as for the homogeneous system, we expand the relative-phase and relative-density fluctuations as
\beq
\hat{\phi}(z,t) =  \sum_{j>0} \hat{\phi}_j(t) f_j(z) \ \ , \ \hat{n}(z,t) =  \sum_{j>0} \hat{n}_j(t) f_j(z).
\label{eq:expansion_trap}
\eeq
The equations of motion read
\begin{eqnarray}
\partial_t\hat{\phi}(z,t) = -\pi v_N \hat{n}(z,t) \nonumber \\
\pi\partial_t\hat{n}(z,t) = -\partial_z [v_J(z) \partial_z \hat{\phi}(z,t)].
\end{eqnarray}
Inserting the expansion \eqref{eq:expansion_trap} and using the expression of $n_0(z)$ in the Thomas-Fermi regime, we find that the eigenfunctions of the problem satisfy the equation 
\beq
(1-x^2)f_j^{\prime \prime}(x) - 2xf_j^\prime(x)+\frac{2\omega_j^2}{\omega^2}f_j(x)=0
\eeq
where  $x=z/R$ and $f_j^\prime$ (resp. $f_j^{\prime \prime}$) denotes the first (resp. second) spatial derivative of $f_j$  with respect to $x$.
The solutions are well known and can be expressed in terms of Legendre polynomials $P_j$:  
\beq
f_j(z)=\sqrt{j+\frac{1}{2}} \ P_j(x) \ \text{with} \ x=z/R \ \ , \ \omega_j=\omega\sqrt{j(j+1)/2}.
\label{eq:solutions}
\eeq

The initial conditions for the mode occupation numbers after splitting for the trapped system are \textit{a priori} different than for the homogeneous system (Eq. \eqref{eq:init_conditions}). This is due to the fact that the amount of noise is proportional to the local density of atoms and is thus lower at the edges of the cloud than in the center. 
In the quantum optics picture, this would be the analogue of a local beam splitter.
More specifically, the fluctuations are given by
\begin{eqnarray}
\langle\hat{n}(z)\hat{n}(z^\prime)\rangle |_{t=0} = \frac{\xi_n^2(z) n_0(z)}{2}\delta(z-z^\prime) \nonumber \\
\\
\langle\hat{\phi}(z)\hat{\phi}(z^\prime)\rangle |_{t=0} = \frac{1}{2\xi_n^2(z) n_0(z)}\delta(z-z^\prime) \nonumber ,
\end{eqnarray}
where we introduced the local squeezing factor $\xi_n^2(z)$.
It can be shown  that the corresponding mode occupation numbers $\langle \hat{n}_j^\dagger \hat{n}_j \rangle|_{t=0}$ for the trapped system are close to that for the homogeneous system (see Eq.~\eqref{eq:init_conditions}) and only differ slightly for the first two modes $j=1,2$ \cite{noteInitCondTrap}. 
As this small difference does not influence the final result, we will assume for clarity the same occupation for all modes, with initial density fluctuations given by 
\beq
\langle \hat{n}_j^\dagger \hat{n}_j \rangle|_{t=0}=\frac{\npeak}{2R}.
\label{eq:initi_cond_trap}
\eeq 
With this choice of initial conditions, the dephased state has the same effective temperature $T\stxt{eff}=\npeak g/2 k\stxt{B}$ as for the homogeneous system. 
These initial conditions correspond to mode occupation numbers $\langle\hat{b}_j^\dagger \hat{b}_j\rangle = k\stxt{B}T\stxt{eff}/\hbar\omega_j$ which are the same as in thermal equilibrium \cite{Petrov2004} but with an effective temperature $T\stxt{eff}$.

\begin{figure}[h!]
	\centering
	\includegraphics[width=0.7\textwidth]{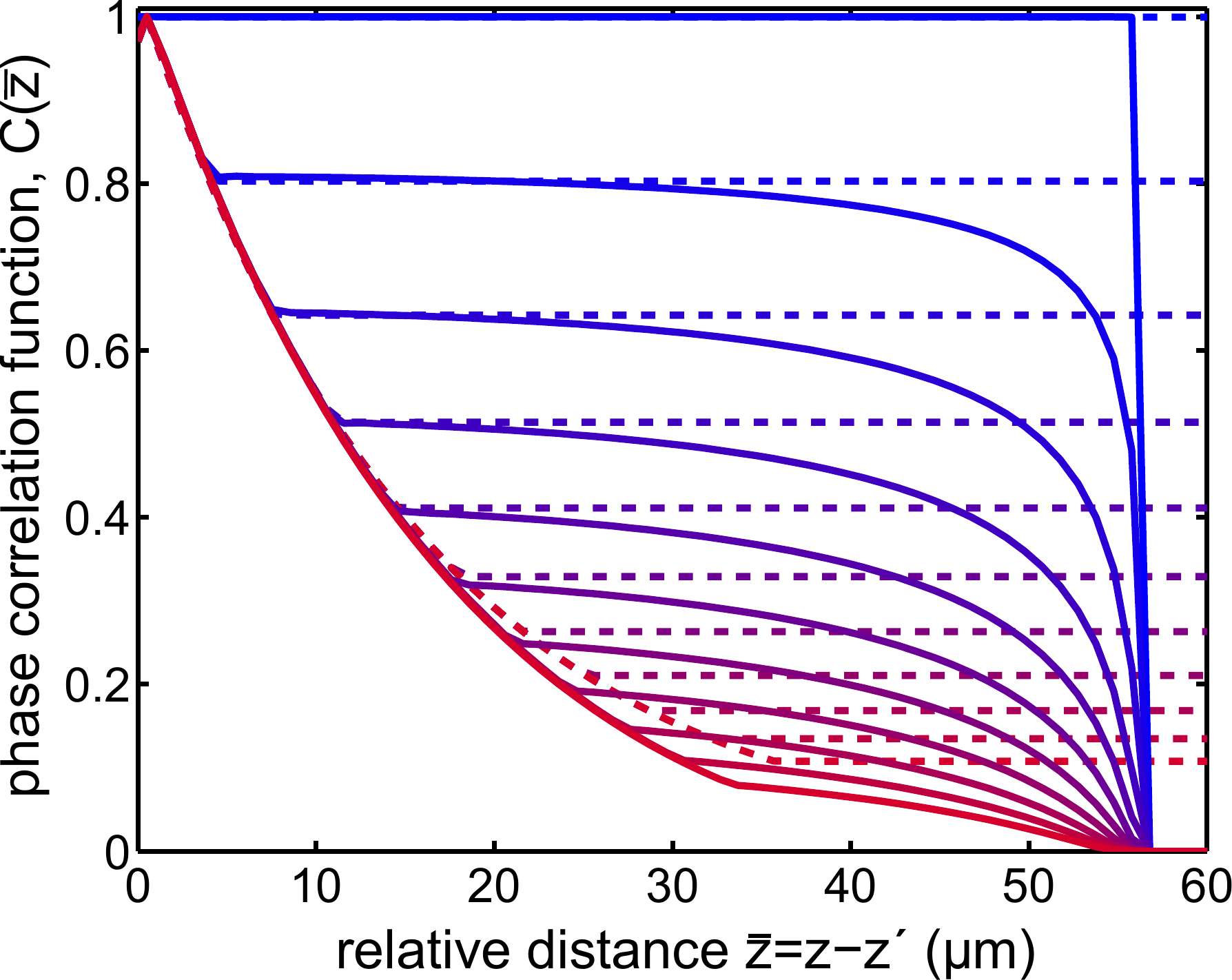} 
	\caption{Evolution of the relative phase correlation function for the trapped system (solid) and for the homogeneous system (dashed). We plot the correlation function $C(z,z^\prime)$ with one point at the center of the cloud ($z^\prime=0$, $z=\bar z$). The parameters are $\omega_\perp=2\pi\times 1400 \ \text{Hz}$ and $\omega=2\pi\times 7 \ \text{Hz}$ and we chose the same peak density $\npeak=46 \ \text{atoms}/\text{micron}$ for both systems, corresponding to $7000$ atoms \cite{Petrov2000}. The evolution times are the same as in Fig.~\ref{fig:time_evolution} (0 to 10 ms from top to bottom). Note that the correlation function goes to zero at the Thomas-Fermi radius for the trapped system.}
	\label{fig:comparison_homog_trap}
\end{figure}

Finally, combining Eqs. \eqref{eq:solutions} and \eqref{eq:initi_cond_trap}  we obtain for the two-point phase variance:
\begin{align}
	\langle\Delta\phi_{zz'}(t)^2\rangle =  \frac{n_0 \pi^2 v_N^2 }{2 R} \sum_{j=1}^\infty \frac{\sin(\omega_j t)^2}{\omega_j^2}\left[f_j(z)-f_j(z')\right]^2. 
	\label{eq:trap}
\end{align}
The expression has a similar form as the one  found for the homogeneous system (equation \eqref{eq:phase_variance}). Again, the time dependent term denotes the oscillation of energy between density and phase fluctuations, and the position-dependent term reflects the spatial phase fluctuations. 

The comparison of the two-point relative-phase correlation functions for the trapped and homogeneous system is shown in Fig.~\ref{fig:comparison_homog_trap}, for the same value of the peak density $\npeak$. 
The general behaviour is similar to the homogeneous system, with a correlation function being exponential up to a characteristic distance beyond which partial long-range order remains. However, three differences appear between the two systems. First, the effective correlation length $\lambda\stxt{eff}$ corresponding to the $1/e$ value of $C(\bar z)$ is slightly lower than in the homogeneous case (see Eq. \eqref{eq:condition_multimode}), due to the presence of the Legendre polynomials in the sum \eqref{eq:trap}. Second,  the long-range order plateaus appear at slightly shorter separations because the effective correlation length is lower, and because of a smaller velocity of correlations, as we will see in the next paragraph. Finally, the correlation function drops to zero at the edge of the cloud corresponding here to $R\approx 56 \ \text{micron}$.

\subsection{Light-cone effect}

\begin{figure}[h!]
	\centering
	\includegraphics[width=0.7\textwidth]{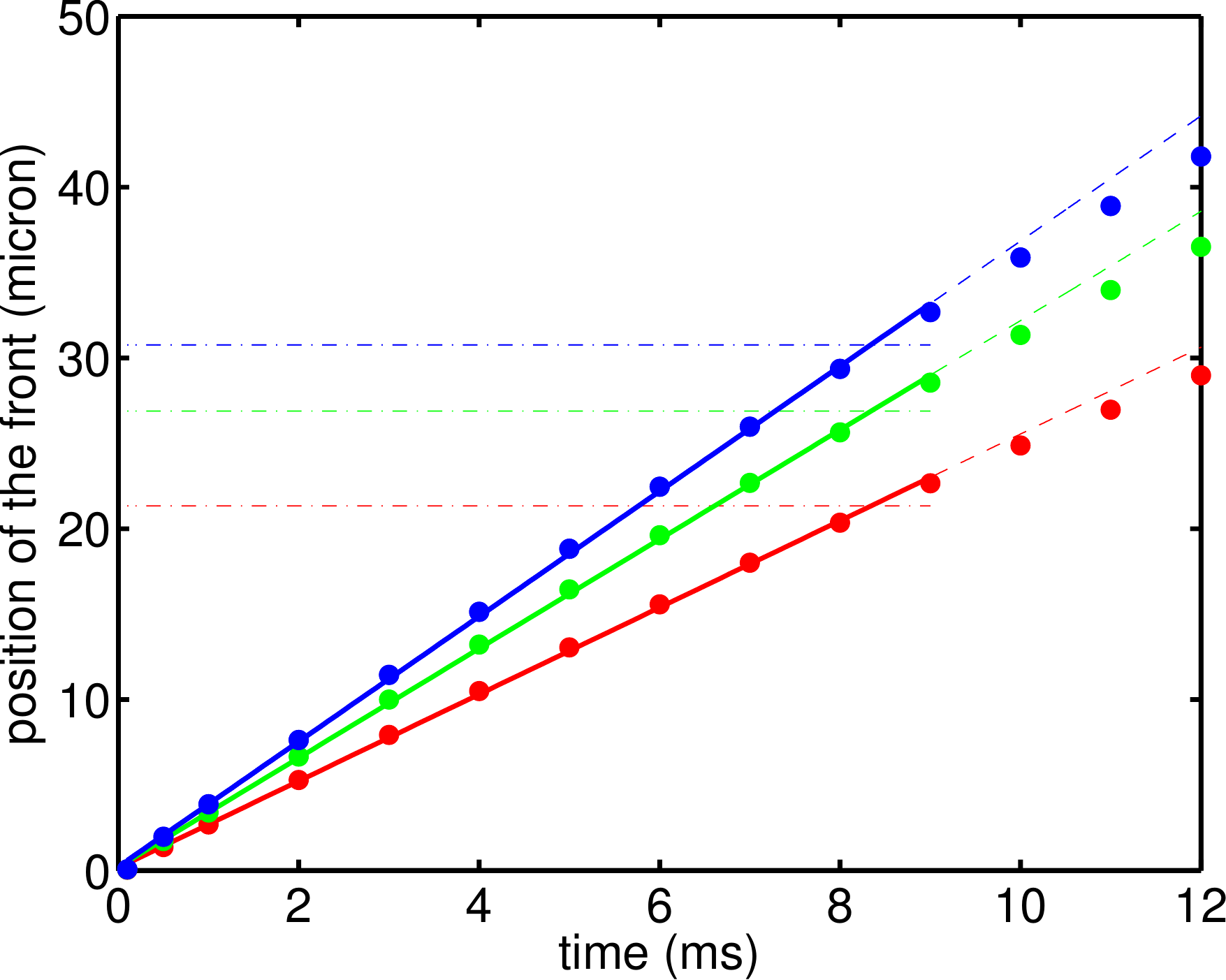} 
	\caption{Position of the front of correlation for the trapped system  for different atom numbers: 3000 (red), 6000 (green) and 9000 (blue); the radial and longitudinal trap frequencies are $\omega_\perp/2\pi= 1400 \ \text{Hz}$ and $\omega/2\pi= 7 \ \text{Hz}$, respectively. The solid line is a linear fit for times $< 10 \ \text{ms}$ to extract the characteristic velocity. The horizontal dot-dashed lines indicate half the Thomas-Fermi radius, $R/2$.}
	\label{fig:light_cone_trap}
\end{figure}

In the homogeneous case, the light-cone-like dynamics of correlations directly comes from the Lorentz invariant form of the equation for the relative phase field. Such a light-cone condition is not obvious for the trapped system, where the speed of sound depends on the local density in the gas. Numerically computing the second derivative of the phase variance from eq. \eqref{eq:trap}, we find a well-defined front of correlations, the position of which linearly evolves in time. 
From this, we determine the characteristic velocity by extracting the position of the correlation front for various evolution times. 
The procedure is illustrated for three different atom numbers (peak densities) in Fig.~\ref{fig:light_cone_trap}. 

For evolution times corresponding to a distance travelled by the quasi-particles approaching the edge of the cloud at $\pm R$, the dynamics is dominated by the finite size effect (bending of the correlation function, see Fig. \ref{fig:comparison_homog_trap}) and the position of the front slightly deviates from the linear scaling in time. This is illustrated by indicating the separation corresponding to half the Thomas-Fermi radius in Fig.~\ref{fig:light_cone_trap} (horizontal lines). However, exploring this region experimentally would be very hard as it would require a very high number of realizations in order to obtain a smooth correlation function in that range of separations (more than $10^4$ realizations typically \cite{Langen2013}).

\begin{figure}[h!]
	\centering
	\includegraphics[width=0.7\textwidth]{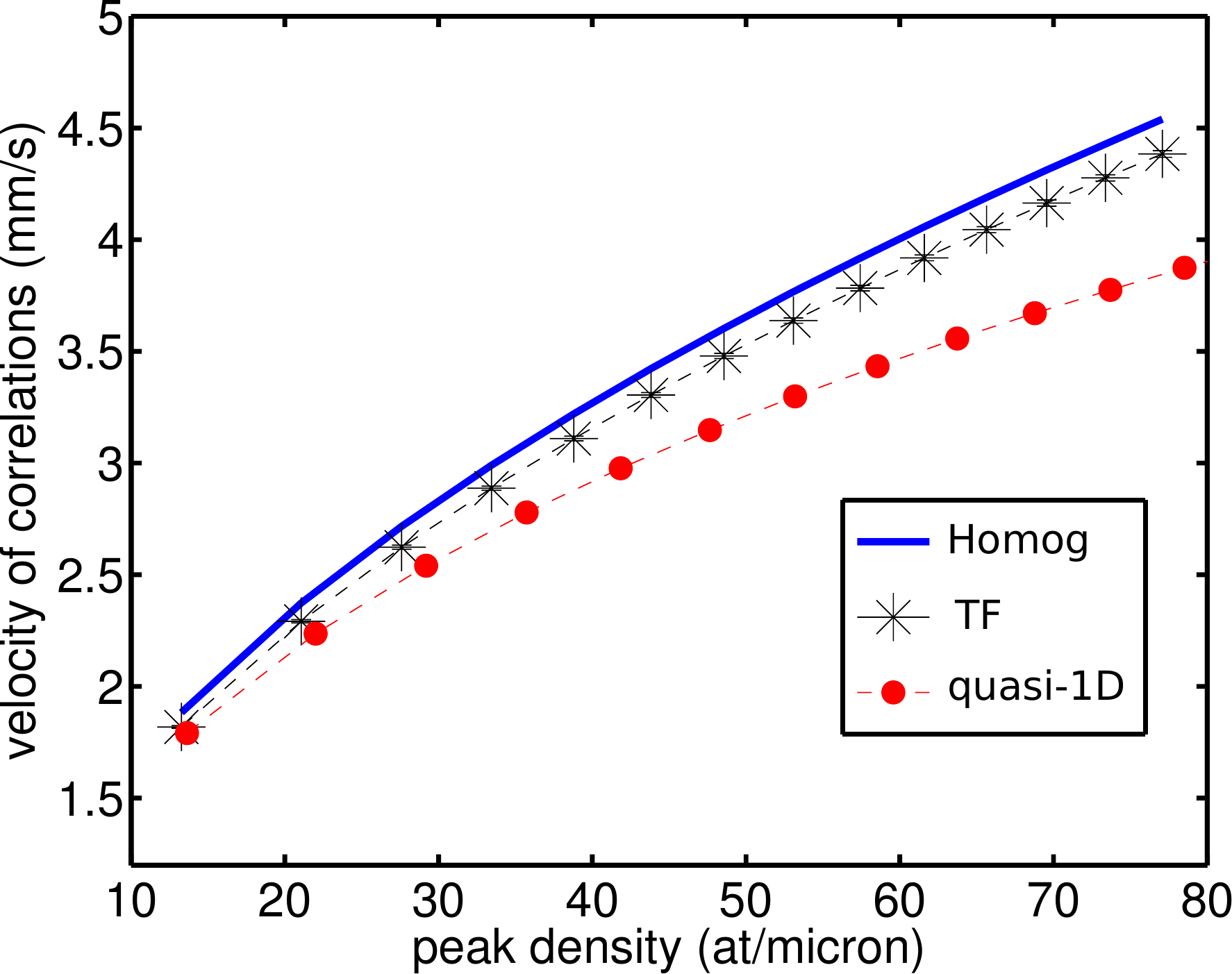} 
	\caption{Comparison of the velocity of correlations for the homogeneous system, the Thomas-Fermi regime (TF), and the quasi-1D regime.  }
	\label{fig:comparison_velocities}
\end{figure}

In typical experiments, the gases are rather in the quasi-1D regime than perfectly 1D (i.e. $ \mu , \ k\stxt{B} T\le \hbar\omega_\perp$). The effect of the radial extension of the cloud can be integrated out, leading to a  density profile in the longitudinal direction that is modified with respect to the Thomas-Fermi (TF) regime \cite{Gerbier2004}. 
Treating  the density profile as an inverted parabola with a modified peak density and an effective radius, we can use Eq. \eqref{eq:trap} to compute the dynamics of phase fluctuations and determine the position of the front of correlations numerically in the quasi-1D regime. The results are presented in 
Fig. \ref{fig:comparison_velocities} and show a lower value for the speed of correlations  ($\sim 10\%$) for the same atom number,  due to the slightly lower value of the radius ($\sim 4\%$) in the quasi-1D regime than in the purely 1D Thomas-Fermi regime.
We emphasize that the lower velocity found for the quasi-1D system than for the TF cloud is due to the more pronounced finite size effect captured in the Legendre polynomials $P_j(z/R)$ in Eq. \eqref{eq:trap}, and not by the change in density between the two regimes (the latter would rather lead  to a higher velocity of correlations in the quasi-1D regime because $\npeak|_{\text{quasi-1D}}>\npeak|_{\text{TF}}$ by about $10\%$.)
Therefore, this difference could  not have been captured by a homogeneous system calculation, which would have overestimated the velocity of correlations. Our trapped calculation for the quasi-1D regime is in good agreement with experiments \cite{Langen2013}.

\section{Recurrences of coherence}
In the final part of this article, we study the effect of the trapping potential on the recurrences of coherence which are expected to occur when the phase quadrature completes  one oscillation. 
Full recurrences manifest themselves in a phase correlation function that returns to its initial value $C(\bar z) = 1$ for all $\bar z$, meaning that the initial state is re-established.
These recurrences of coherence show  similarities with the many-body revivals in optical lattices \cite{Greiner2002,Will2010}, the characterization of which is known to be strongly affected by an external trapping potential \cite{Will2010}.

Moreover, investigating possible  recurrences of coherence is of great importance to characterize the long-time evolution of the system in detail, where physics beyond the harmonic approximation is expected to lead to thermalization. Possible mechanisms leading to this thermalization may be described, for example, by non-linear terms present in the 1D Hamiltonian and effectively leading to phonon-phonon scattering \cite{Mazets2009} or by integrability breaking terms due the non-perfect 1D-ness of the system  \cite{Mazets2008,Tan2010}. 
Understanding the exact form and the scaling of these recurrences with the system parameters in the harmonic approximation is thus important to distinguish genuine many-body effects (quasi-particle relaxation) from the integrable dynamics.

\begin{figure}[h!]
	\centering
	\includegraphics[width=0.9\textwidth]{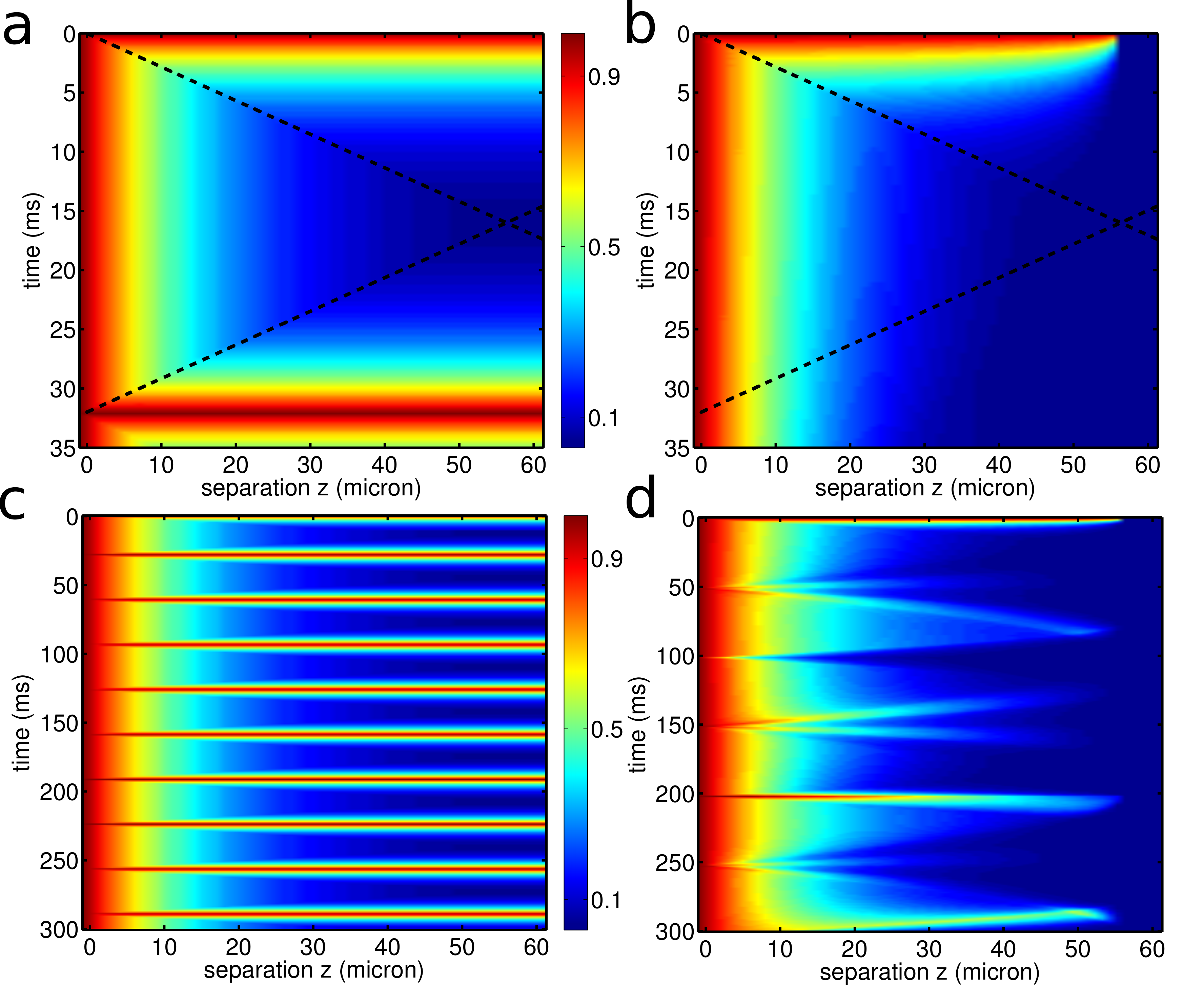} 
	\caption{Time evolution of the relative-phase correlation function for the homogeneous (left, a and c) and trapped (right, b and d) systems. The color-scale indicates the degree of correlation (red: high correlation, blue: low correlation). The top row illustrates the relaxation to the prethermalized state with the speed of sound as characteristic velocity for the decay of correlations (dahsed line is $\bar z =2ct$).
	 In the homogeneous case, the initial state is re-established at times which are multiples of the system length divided by the characteristic velocity. 
	  In the trapped case, the recurrences are only partial and the more complex structure is due to the incommensurate ratios of the mode frequencies $\omega_j$. In this time window (0-300 ms), the strongest recurrence is observed at $202 \ \text{ms}$ ($\omega/2\pi=7 \ \text{Hz}$). }
	\label{fig:revival_2D_corrFunc}
\end{figure}

In the purely harmonic approximation, the energy initially introduced by the splitting process in the density quadrature oscillates in time between the relative phase and relative density fluctuations. In the homogeneous system, this corresponds to full recurrences of phase coherence at times $t\stxt{rev}=\mathcal{L}/2c$, where the value of the two-point phase correlation function comes back to 1, i.e. where the initial state is re-established, see Fig.~\ref{fig:revival_2D_corrFunc}.
The recurrences of coherence emerge in the form of an inverse light-cone-like evolution, symmetric to the  dephasing leading to the prethermalized state. This is illustrated by the dashed black lines in Fig.~\ref{fig:revival_2D_corrFunc} (a), which shows the position of the front where long-range phase coherence starts.

\begin{figure}[h!]
	\centering
	\includegraphics[width=0.6\textwidth]{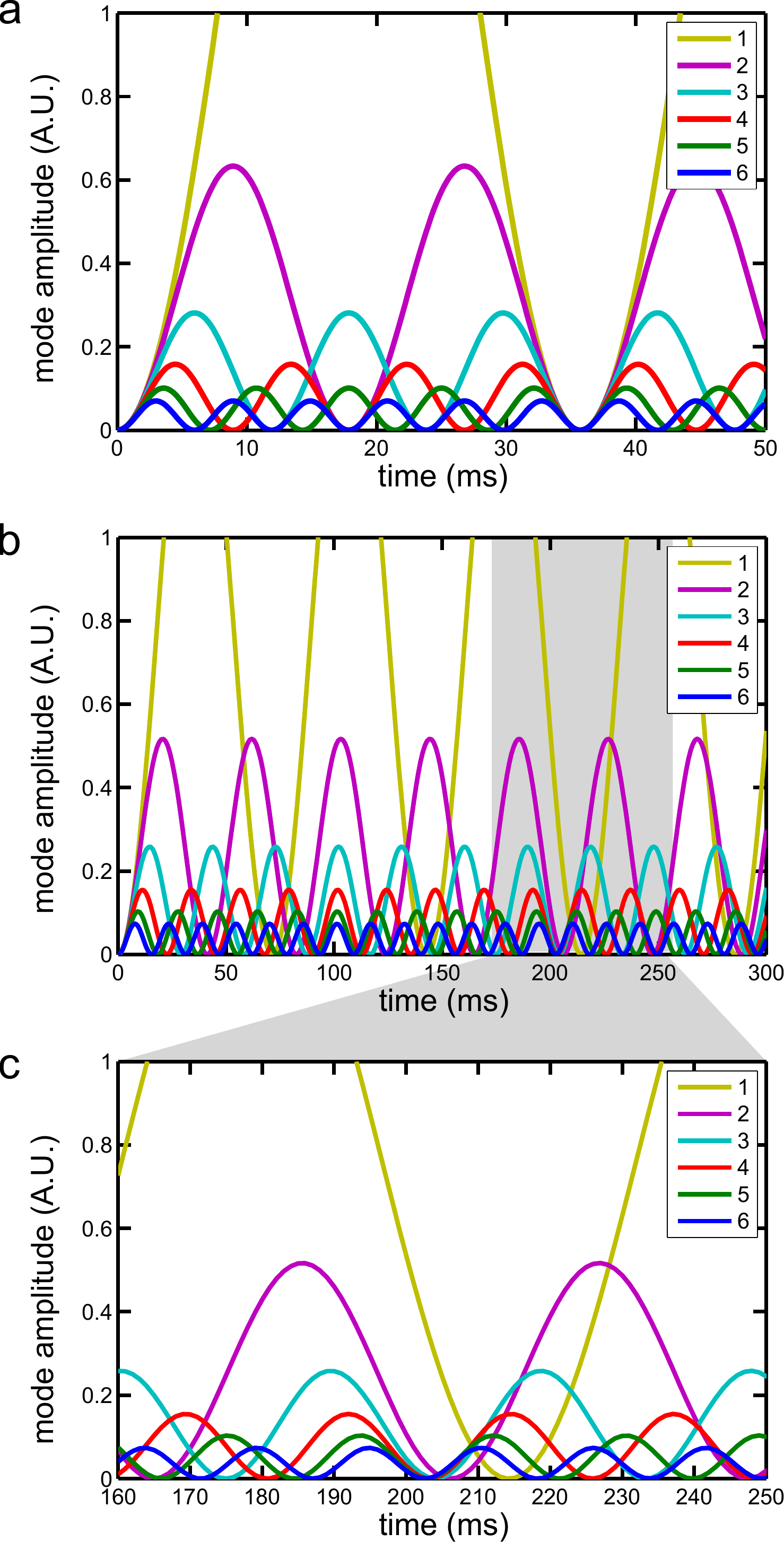} 
	\caption{Visualization of the first six mode amplitudes $(\sin(\omega_j t)/\omega_j)^2$ contributing to the phase variance : (a) homogeneous system, (b) trapped system. For the trap system, the strongest recurrence of coherence appears when most of the modes complete their oscillation at about the same time, which occurs around 202 ms for the parameters of Fig.~\ref{fig:revival_2D_corrFunc} where $\omega/2\pi = 7 \ \text{Hz}$. (c) Zoom around the recurrence time.}
	\label{fig:visualize_modes_trap}
\end{figure}

In the trapped system, the incommensurate ratios between the excitation frequencies $\omega_j$ lead to partial recurrences at times which strongly differ from the ones in the homogeneous case (see Fig.~\ref{fig:revival_2D_corrFunc} b and d). Contrary to the homogeneous case, no recurrence is observed at early times (compare (a) and (b)).
In contrast, the strongest recurrence is observed at much longer times, where the interference of the different modes  is the most favourable (at $202 \ \text{ms}$ for the parameters of Fig.~\ref{fig:revival_2D_corrFunc}). 
In that case, visualizing the different mode amplitudes in Fig.~\ref{fig:visualize_modes_trap} reveals that this recurrence corresponds to a point in time slightly before the lowest $j=1$ mode completes its third oscillation (at time $t=3\pi/\omega_1\approx 71 \ \text{ms}$ for $\omega=2\pi\times 7 \ \text{Hz}$), while the second $j=2$  mode almost completes its fifth oscillation.
Even at longer evolution times, full recurrence ($C(\bar z)=1$) cannot be observed for the trap system because of the incommensurate ratio of the mode frequencies $\omega_j$.

\begin{figure}[h!]
	\centering
	\includegraphics[width=0.8\textwidth]{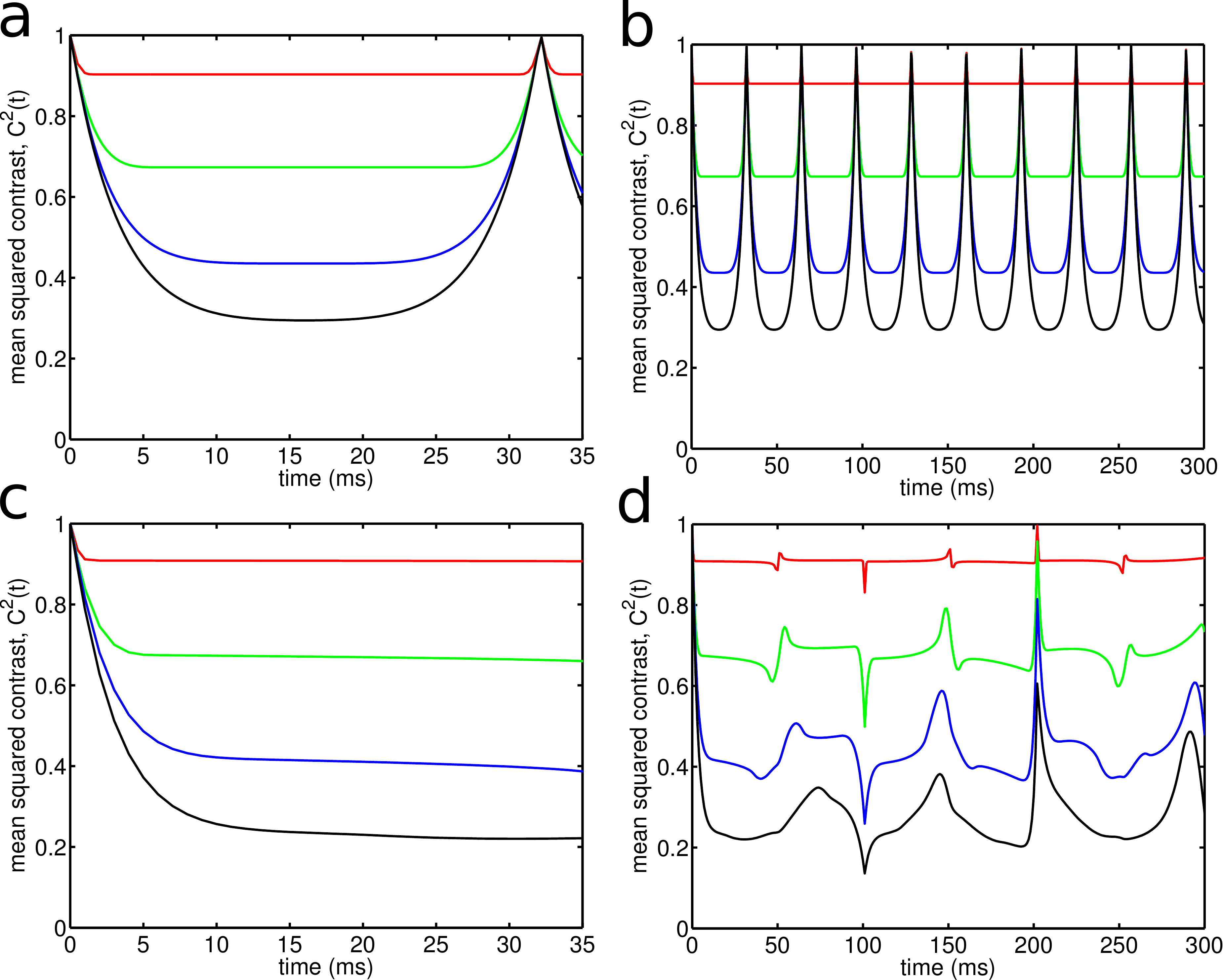} 
	\caption{Time-evolution of the integrated squared contrast showing the recurrences of coherence.  The different traces correspond to different integration length ($L= 5, \ 20, \ 50, \ 90 \ \text{micron}$ from top to bottom ). Top: homogeneous system, bottom: trapped system. The left column shows the dephasing at short times to the prethermalized state, as investigated in \cite{Kuhnert2013}. The right column shows the long time evolution with the strong differences for the recurrences of coherence between the homogeneous and the trap systems.}
	\label{fig:contrast_evolution}
\end{figure}

For a more intuitive illustration of the recurrences, we show in Fig.~\ref{fig:contrast_evolution} the time evolution of the mean squared contrast (integrated over a region of length $L$), $C^2(t)$,   which is a simple measure of coherence in the system \cite{Kuhnert2013}. It can be calculated from a double integration of the PCF, and is directly accessible in experiments from the interference patterns integrated over a size $L$. 
Here again, we observe the more complex structure of the dynamics due to the trap, with  a clear shift of the recurrence time with respect to the homogeneous case, and a different form of the recurrence.

\section{Conclusion.}
\label{par:conclusion}

We have theoretically studied the dynamics of a coherently split 1D Bose gas in the harmonic approximation, considering phononic excitations and taking into account a longitudinal trapping potential. 
We showed that the relaxation to the prethermalized dephased state is local and that the dephasing spreads in time through the system in a light-cone-like evolution with a characteristic velocity. 
Including the trapping potential in the model, we showed that the trap has small effects at short evolution times, but important effects at long evolution times (compared to the dephasing time-scale). More precisely, the trapping potential does not affect the nature of the dephasing which is still local, but affects the velocity of correlations.
At longer evolution  times, the trap has an important effect on the recurrences of coherence due to the rephasing of elementary excitations in a finite size system.
In our present work we restricted ourselves to the Luttinger liquid model, which treats the phonons as non interacting quasi-particles. At long times, much longer than the dephasing time, we expect that physics beyond the Luttinger liquid approximation, such as  scattering or decay of phonons, will emerge, which will lead to further relaxation and thermalization \cite{Aarts2001,Gasenzer2005,Grisins2011,Mazets2011PRA}. Also in this case it will be important to take the trapping potential into account when comparing experimental data to the theoretical models.




\section*{Acknowledgements.}
\label{par:acknowledgements}
We thank W. Rohringer, B. Rauer, M. Kuhnert, T. Schweigler, V. Kasper, S. Erne and J.-F. Schaff for discussions. This work was supported by the Austrian Science Fund FWF through the Wittgenstein Prize, Lise Meitner grant M-1423 for R.G., the COQUS doctoral school W1210 for T.L., project P22590-N16 for I.E.M., and by the EU (ERC advance grant number Quantum Relax).

\section*{References}

\bibliography{theory_paper}

\begin{thebibliography}{10}

\bibitem{CazalillaRigol2010NJP}
M~A Cazalilla and M~Rigol.
\newblock Focus on dynamics and thermalization in isolated quantum many-body
  systems.
\newblock {\em New Journal of Physics}, 12(5):055006, 2010.

\bibitem{Polkovnikov2011}
Anatoli Polkovnikov, Krishnendu Sengupta, Alessandro Silva, and Mukund
  Vengalattore.
\newblock \textit{Colloquium} : Nonequilibrium dynamics of closed interacting
  quantum systems.
\newblock {\em Rev. Mod. Phys.}, 83:863--883, Aug 2011.

\bibitem{Neumann1929}
J~v Neumann.
\newblock Beweis des ergodensatzes und des h-theorems in der neuen mechanik.
\newblock {\em Zeitschrift f{\"u}r Physik}, 57(1-2):30--70, 1929.

\bibitem{Srednicki1994}
Mark Srednicki.
\newblock Chaos and quantum thermalization.
\newblock {\em Phys. Rev. E}, 50:888--901, Aug 1994.

\bibitem{Rigol2008ETH}
Marcos Rigol, Vanja Dunjko and Maxim Olshanii, Thermalization and its mechanism
  for generic isolated quantum systems, Nature 452, 854 (2008).

\bibitem{Berges2004}
J.~Berges, Sz. Bors\'anyi, and C.~Wetterich.
\newblock Prethermalization.
\newblock {\em Phys. Rev. Lett.}, 93:142002, 2004.

\bibitem{Rigol2007PRL}
Marcos Rigol, Vanja Dunjko, Vladimir Yurovsky, and Maxim Olshanii.
\newblock Relaxation in a completely integrable many-body quantum system: An
  \textit{Ab~Initio} study of the dynamics of the highly excited states of 1d
  lattice hard-core bosons.
\newblock {\em Phys. Rev. Lett.}, 98:050405, Feb 2007.

\bibitem{Berges2008}
J\"urgen Berges, Alexander Rothkopf, and Jonas Schmidt.
\newblock Nonthermal fixed points: Effective weak coupling for strongly
  correlated systems far from equilibrium.
\newblock {\em Phys. Rev. Lett.}, 101:041603, Jul 2008.

\bibitem{Eckstein2009}
Martin Eckstein, Marcus Kollar, and Philipp Werner.
\newblock Thermalization after an interaction quench in the hubbard model.
\newblock {\em Phys. Rev. Lett.}, 103:056403, 2009.

\bibitem{Kollar2011}
Marcus Kollar, F.~Alexander Wolf, and Martin Eckstein.
\newblock Generalized gibbs ensemble prediction of prethermalization plateaus
  and their relation to nonthermal steady states in integrable systems.
\newblock {\em Phys. Rev. B}, 84:054304, 2011.

\bibitem{Nowak2013}
Non-thermal fixed points: universality, topology, \& turbulence in Bose gases,
  Boris Nowak, Sebastian Erne, Markus Karl, Jan Schole, Dénes Sexty, Thomas
  Gasenzer, arXiv:1302.1448 (2013).

\bibitem{Bloch2008}
Immanuel Bloch, Jean Dalibard, and Wilhelm Zwerger.
\newblock Many-body physics with ultracold gases.
\newblock {\em Rev. Mod. Phys.}, 80:885--964, Jul 2008.

\bibitem{Cazalilla2011RevModPhys}
M.~A. Cazalilla, R.~Citro, T.~Giamarchi, E.~Orignac, and M.~Rigol.
\newblock One dimensional bosons: From condensed matter systems to ultracold
  gases.
\newblock {\em Rev. Mod. Phys.}, 83:1405--1466, Dec 2011.

\bibitem{Gring2012}
M.~Gring, M.~Kuhnert, T.~Langen, T.~Kitagawa, B.~Rauer, M.~Schreitl, I.~Mazets,
  D.~A.~Adu Smith, E.~Demler, and J.~Schmiedmayer.
\newblock Relaxation and prethermalization in an isolated quantum system.
\newblock {\em Science}, 2012.

\bibitem{Gerving2012}
C.S. Gerving, T.M. Hoan, B.J. Land, M.~Anquez, C.D. Hamley, and M.S. Chapman.
\newblock Non-equilibrium dynamics of an unstable quantum pendulum explored in
  a spin-1 bose-einstein condensate.
\newblock {\em Nature Communications}, 3:1169, 2012.

\bibitem{Cheneau2012}
M.~Cheneau, P.~Barmettler, D.~Poletti, M.~Endres, P.~Schau\ss, T.~Fukuhara,
  C.~Gross, I.~Bloch, C.~Kollath, and S.~Kuhr.
\newblock Light-cone-like spreading of correlations in a quantum many-body
  system.
\newblock {\em Nature}, 481:484, 2012.

\bibitem{Hung2013}
Chen-Lung Hung, Victor Gurarie, Cheng Chin, From Cosmology to Cold Atoms:
  Observation of Sakharov Oscillations in a Quenched Atomic Superfluid, Science
  341, p. 1213-1215 (2013).

\bibitem{Sagi2012}
Yoav Sagi, Tara~E. Drake, Rabin Paudel, and Deborah~S. Jin.
\newblock Measurement of the homogeneous contact of a unitary fermi gas.
\newblock {\em Phys. Rev. Lett.}, 109:220402, Nov 2012.

\bibitem{Schmidutz2013}
Tobias F. Schmidutz, Igor Gotlibovych, Alexander L. Gaunt, Robert P. Smith, Nir
  Navon, Zoran Hadzibabic, Quantum Joule-Thomson Effect in a Saturated
  Homogeneous Bose Gas, arXiv:1309.1441 (2013).

\bibitem{Kuhnert2013}
M.~Kuhnert, R.~Geiger, T.~Langen, M.~Gring, B.~Rauer, T.~Kitagawa, E.~Demler,
  D.~Adu~Smith, and J.~Schmiedmayer.
\newblock Multimode dynamics and emergence of a characteristic length scale in
  a one-dimensional quantum system.
\newblock {\em Phys. Rev. Lett.}, 110:090405, Feb 2013.

\bibitem{Langen2013}
Tim Langen, Remi Geiger, Maximilian Kuhnert, Bernhard Rauer, Joerg
  Schmiedmayer, Local emergence of thermal correlations in an isolated quantum
  many-body system, Nature Physics 9, 640 (2013),.

\bibitem{Bistritzer2007}
R~Bistritzer and E~Altman.
\newblock {Intrinsic dephasing in one-dimensional ultracold atom
  interferometers.}
\newblock {\em PNAS}, 104(24):9955--9, June 2007.

\bibitem{Kitagawa2011}
Takuya Kitagawa, Adilet Imambekov, J\"org Schmiedmayer, and Eugene Demler.
\newblock The dynamics and prethermalization of one dimensional quantum systems
  probed through the full distributions of quantum noise.
\newblock {\em New J. Phys.}, 13:073018, 2011.

\bibitem{Burkov2007}
A.~A. Burkov, M.~D. Lukin, and E.~Demler.
\newblock Decoherence dynamics in low-dimensional cold atom interferometers.
\newblock {\em Phys. Rev. Lett.}, 98(20):200404, May 2007.

\bibitem{Stimming2011}
H.-P. Stimming, N.~J. Mauser, J.~Schmiedmayer, and I.~E. Mazets.
\newblock Dephasing in coherently split quasicondensates.
\newblock {\em Phys. Rev. A}, 83(2):023618, Feb 2011.

\bibitem{Mazets2009}
I.~E. {Mazets} and J.~{Schmiedmayer}.
\newblock {Dephasing in two decoupled one-dimensional Bose-Einstein condensates
  and the subexponential decay of the interwell coherence}.
\newblock {\em European Physical Journal B}, 68:335--339, 2009.

\bibitem{Petrov2000}
D.~S. Petrov, G.~V. Shlyapnikov, and J.~T.~M. Walraven.
\newblock Regimes of quantum degeneracy in trapped 1d gases.
\newblock {\em Phys. Rev. Lett.}, 85(18):3745--3749, Oct 2000.

\bibitem{Schumm2005}
T.~Schumm, S.~Hofferberth, L.~M. Andersson, S.~Wildermuth, S.~Groth,
  I.~Bar-Joseph, J.~Schmiedmayer, and P.~Kruger.
\newblock {Matter-wave interferometry in a double well on an atom chip}.
\newblock {\em Nature Physics}, 1(1):57--62, 2005.

\bibitem{Olshanii1998}
M.~Olshanii.
\newblock Atomic scattering in the presence of an external confinement and a
  gas of impenetrable bosons.
\newblock {\em Phys. Rev. Lett.}, 81(5):938--941, Aug 1998.

\bibitem{Mora2003}
Christophe Mora and Yvan Castin.
\newblock Extension of bogoliubov theory to quasicondensates.
\newblock {\em Phys. Rev. A}, 67:053615, May 2003.

\bibitem{Langen2013epjst}
Tim Langen, Michael Gring, Maximilian Kuhnert, Bernhard Rauer, Remi Geiger,
  DavidAdu Smith, IgorE. Mazets, and Jörg Schmiedmayer.
\newblock Prethermalization in one-dimensional bose gases: Description by a
  stochastic ornstein-uhlenbeck process.
\newblock {\em The European Physical Journal Special Topics}, 217(1):43--53,
  2013.

\bibitem{Bouchoule2003}
Nicholas~K. Whitlock and Isabelle Bouchoule.
\newblock Relative phase fluctuations of two coupled one-dimensional
  condensates.
\newblock {\em Phys. Rev. A}, 68:053609, 2003.

\bibitem{Giamarchi2004}
Thierry Giamarchi.
\newblock {\em Quantum physics in one dimension}.
\newblock Internat. Ser. Mono. Phys. Clarendon Press, Oxford, 2004.

\bibitem{Castin1997}
Yvan Castin and Jean Dalibard.
\newblock Relative phase of two bose-einstein condensates.
\newblock {\em Phys. Rev. A}, 55:4330--4337, Jun 1997.

\bibitem{Javanainen1997}
Juha Javanainen and Martin Wilkens.
\newblock Phase and phase diffusion of a split bose-einstein condensate.
\newblock {\em Phys. Rev. Lett.}, 78:4675--4678, Jun 1997.

\bibitem{Leggett1998}
A.~J. Leggett and F.~Sols.
\newblock Comment on ``phase and phase diffusion of a split bose-einstein
  condensate''.
\newblock {\em Phys. Rev. Lett.}, 81:1344--1344, Aug 1998.

\bibitem{Grond2010}
Julian Grond, Ulrich Hohenester, Igor Mazets, and J\"org Schmiedmayer.
\newblock Atom interferometry with trapped bose-einstein condensates: impact of
  atom-atom interactions.
\newblock {\em New Journal of Physics}, 12(6):065036, 2010.

\bibitem{Maussang2010}
Kenneth Maussang, G.~Edward Marti, Tobias Schneider, Philipp Treutlein, Yun Li,
  Alice Sinatra, Romain Long, J\'er\^ome Est\`eve, and Jakob Reichel.
\newblock Enhanced and reduced atom number fluctuations in a bec splitter.
\newblock {\em Phys. Rev. Lett.}, 105:080403, Aug 2010.

\bibitem{Berrada2013}
T. Berrada, S. van Frank, R. B\"ucker, T. Schumm, J.-F. Schaff, J.
  Schmiedmayer, Integrated Mach-Zehnder interferometer for Bose-Einstein
  condensates, Nature Communications 4, 2077 (2013).

\bibitem{Calabrese2006}
Pasquale Calabrese and John Cardy.
\newblock Time dependence of correlation functions following a quantum quench.
\newblock {\em Phys. Rev. Lett.}, 96:136801, Apr 2006.

\bibitem{Mitra2011}
Aditi Mitra and Thierry Giamarchi.
\newblock Mode-coupling-induced dissipative and thermal effects at long times
  after a quantum quench.
\newblock {\em Phys. Rev. Lett.}, 107:150602, Oct 2011.

\bibitem{Barmettler2012}
Peter Barmettler, Dario Poletti, Marc Cheneau, and Corinna Kollath.
\newblock Propagation front of correlations in an interacting bose gas.
\newblock {\em Phys. Rev. A}, 85:053625, May 2012.

\bibitem{Mitra2013}
Aditi Mitra.
\newblock Correlation functions in the prethermalized regime after a quantum
  quench of a spin chain.
\newblock {\em Phys. Rev. B}, 87:205109, May 2013.

\bibitem{Carleo2013arxiv}
Giuseppe Carleo, Federico Becca, Laurent Sanchez-Palencia, Sandro Sorella,
  Michele Fabrizio, Light-Cone Effect and Supersonic Correlations in Bosonic
  Superfluids, arXiv:1310.2246 (2013).

\bibitem{Deuar2013arxiv}
Piotr Deuar, Magdalena Stobinska, Correlation waves after quantum quenches in
  one- to three-dimensional BECs, arXiv:1310.1301 (2013).

\bibitem{Cramer2008}
M.~Cramer, C.~M. Dawson, J.~Eisert, and T.~J. Osborne.
\newblock Exact relaxation in a class of nonequilibrium quantum lattice
  systems.
\newblock {\em Phys. Rev. Lett.}, 100:030602, Jan 2008.

\bibitem{Petrov2004}
D.S. Petrov, D.M. Gangardt and G.V. Shlyapnikov, Low-dimensional trapped gases,
  J. Phys. IV France 116 (2004) 5-44.

\bibitem{noteInitCondTrap}
S. Erne, T. langen, T. Gasenzer et al., in preparation, Nov. 2013.

\bibitem{Gerbier2004}
F.~{Gerbier}.
\newblock {Quasi-1D Bose-Einstein condensates in the dimensional crossover
  regime}.
\newblock {\em Europhysics Letters}, 66:771--777, 2004.

\bibitem{Greiner2002}
Markus Greiner, Olaf Mandel, Theodor~W. Hansch, and Immanuel Bloch.
\newblock Collapse and revival of the matter wave field of a bose-einstein
  condensate.
\newblock {\em Nature}, 419:51--54, 2002.

\bibitem{Will2010}
Sebastian Will, Thorsten Best, Ulrich Schneider, Lucia Hackermüller, Dirk-Sören
  Lühmann and Immanuel Bloch, Time-resolved observation of coherent multi-body
  interactions in quantum phase revivals, Nature 465, 197-201 (2010).

\bibitem{Mazets2008}
I.~E. Mazets, T.~Schumm, and J.~Schmiedmayer.
\newblock Breakdown of integrability in a quasi-1d ultracold bosonic gas.
\newblock {\em Phys. Rev. Lett.}, 100(21):210403, May 2008.

\bibitem{Tan2010}
Shina Tan, Michael Pustilnik, and Leonid~I. Glazman.
\newblock Relaxation of a high-energy quasiparticle in a one-dimensional bose
  gas.
\newblock {\em Phys. Rev. Lett.}, 105:090404, Aug 2010.

\bibitem{Aarts2001}
Gert Aarts and J\"urgen Berges.
\newblock Nonequilibrium time evolution of the spectral function in quantum
  field theory.
\newblock {\em Phys. Rev. D}, 64:105010, Oct 2001.

\bibitem{Gasenzer2005}
Thomas Gasenzer, J\"urgen Berges, Michael~G. Schmidt, and Marcos Seco.
\newblock Nonperturbative dynamical many-body theory of a bose-einstein
  condensate.
\newblock {\em Phys. Rev. A}, 72:063604, Dec 2005.

\bibitem{Grisins2011}
Pjotrs Grisins and Igor~E. Mazets.
\newblock Thermalization in a one-dimensional integrable system.
\newblock {\em Phys. Rev. A}, 84:053635, Nov 2011.

\bibitem{Mazets2011PRA}
I.~E. Mazets.
\newblock Dynamics and kinetics of quasiparticle decay in a
  nearly-one-dimensional degenerate bose gas.
\newblock {\em Phys. Rev. A}, 83:043625, Apr 2011.

\end{thebibliography}
\bibliographystyle{unsrt}



\end{document}